\pdfoutput=1

\documentclass[11pt,twoside,a4paper,cmspaper,final,collab]{cms-tdr}
\def\svnVersion{179509}\def\cmsCernNo{2012-294}\def\cmsCernDate{\today}\def\cmsMessage{Submitted to Physics Letters B}

\begin{document}\cmsNoteHeader{EXO-11-025}

\hyphenation{had-ron-i-za-tion}
\hyphenation{cal-or-i-me-ter}
\hyphenation{de-vices}

\RCS$Revision: 178486 $
\RCS$HeadURL: svn+ssh://svn.cern.ch/reps/tdr2/papers/EXO-11-025/trunk/EXO-11-025.tex $
\RCS$Id: EXO-11-025.tex 178486 2013-03-28 09:30:34Z digiovan $
\newlength\cmsFigWidth
\ifthenelse{\boolean{cms@external}}{\setlength\cmsFigWidth{0.98\columnwidth}}{\setlength\cmsFigWidth{0.85\textwidth}}
\ifthenelse{\boolean{cms@external}}{\providecommand{\cmsLeft}{top}}{\providecommand{\cmsLeft}{left}}
\ifthenelse{\boolean{cms@external}}{\providecommand{\cmsRight}{bottom}}{\providecommand{\cmsRight}{right}}
\newcommand{\Mqst}{\ensuremath{M_{\cPq^*}}\xspace}
\newcommand{\invGeV}{\ensuremath{\,\text{Ge\hspace{-.08em}V}^{-1}}\xspace}
\newcommand{\invPtMuMu}{\ensuremath{1/\pt(\Pgm\Pgm)}\xspace}
\newcommand{\ptMuMu}{\ensuremath{\pt(\Pgm\Pgm)}\xspace}
\newcommand{\qStar}{\ensuremath{\cPq^*}\xspace}
\newcommand{\fs}{\ensuremath{ f_{s} } }
\newcommand{\fp}{\ensuremath{ f^\prime } }
\newcommand{\massCut}{\ensuremath{60\GeV < M_{\Pgm\Pgm} < 120\GeV}}
\newcommand\roostats{\texttt{RooStats}\xspace}
\newcommand\acc{\ensuremath{\mathcal{A}}\xspace}
\newcommand{\anaLumi}{5.0\fbinv}
\newcommand{\limitsCIfsZero}{2.22\TeV}
\newcommand{\limitsCIfsOne}{2.15\TeV}
\newcommand{\limitsGIfsOne}{1.94\TeV}
\cmsNoteHeader{EXO-11-025} 
\title{\texorpdfstring{Search for anomalous production of highly boosted Z bosons decaying to $\Pgmp \Pgmm$ in proton-proton collisions at $\sqrt{s} = 7\TeV$}{Search for anomalous production of highly boosted Z bosons decaying to dimuon final state in pp collisions at sqrt(s) = 7 TeV}}

\date{\today}

\abstract{
Results are reported from a search for the anomalous production of highly
boosted Z bosons with large transverse momentum and decaying to $\Pgmp \Pgmm$.
Such Z bosons may be produced in the decays of new heavy particles. The search
uses pp collision data at $\sqrt{s} = 7$\TeV, corresponding to an integrated
luminosity of 5.0\fbinv recorded with the CMS detector. The shape of the
observed transverse momentum distribution of Z bosons is consistent with
standard model expectations. Constraints are obtained on models predicting the
production of excited quarks decaying via electroweak processes.
Assuming a compositeness scale that is equal to the excited quark mass as well
as transition coupling strengths between Z bosons and excited quarks that are
equal to standard model couplings to quarks, masses of excited quarks below
1.94\TeV are excluded at the 95\% confidence level. For excited quark production
via a novel contact interaction, masses below 2.22\TeV are excluded, even if the
excited quarks do not couple to gluons.}

\hypersetup{%
pdfauthor={CMS Collaboration},%
pdftitle={Search for anomalous production of highly boosted Z bosons decaying to dimuon final state in pp collisions at sqrt(s) = 7 TeV},%
pdfsubject={CMS},%
pdfkeywords={CMS, physics, dimuons, new phenomena, excited quarks}%
}

\maketitle 

In the standard model (SM) of particle physics, the transverse momentum (\pt)
spectrum of Z bosons produced in high energy pp collisions is predicted to be a
smoothly falling distribution. A broad range of new physics models such as quark
compositeness~\cite{baur,baur2}, supersymmetry~\cite{Matchev:1999ft},
technicolor~\cite{Foadi:2007ue}, and extensions of the standard model with new
gauge groups~\cite{RevModPhys.56.579} predict decays of heavy particles
involving Z bosons, which would introduce deviations from a smooth distribution
of the Z transverse momentum spectrum.  The Z boson transverse momentum
distribution was measured by ATLAS~\cite{Aad:2011gj} and
CMS~\cite{Chatrchyan:2011wt} in pp collisions at $\sqrt{s}=7\TeV$ using about
40\pbinv of integrated luminosity, showing good agreement with
next-to-next-to-leading order perturbative QCD calculations. This Letter
presents a search for new phenomena exploiting the dimuon transverse momentum
spectrum, \ptMuMu, of Z boson production in pp collisions. The dimuon spectrum
is used to search for a new heavy particle decaying into high momentum Z
bosons. Because the search is inclusive, no constraints are imposed on the
presence of additional particles in the decay of the hypothetical heavy
particle. The final results are interpreted within the framework of specific models of
excited quark production, $\qStar\to
\Pq\Z,\:\Z \to \Pgmp \Pgmm$.

Compositeness models explain the observed mass hierarchy of quarks and leptons
by introducing quark constituents, predicting a multitude of excited fermion
states, including excited quarks~\cite{baur,baur2}. In the effective Lagrangian
describing the gauge mediated transitions of excited fermions, the couplings to
the strong and electroweak sectors are measured in units of the strengths of the
SM gauge couplings $g_s$ (strong coupling), $g=e\sin\theta_\PW$, and
$g'=e\cos\theta_\PW$, where $e$ is the electron charge and $\theta_\PW$ is the weak
mixing angle. Thus, the corresponding strengths for the new interactions are
parametrized in terms of scale factors $f_s, f$, and $f'$ with respect to the SM
couplings.
Moreover, gauge models of excited quarks can be extended with novel four-fermion
contact interaction terms arising from new strong dynamics.  In hadron
collisions, excited quarks are usually sought in the dijet final state. Results
from previous studies of dijet mass and angular distributions are consistent
with QCD predictions
\cite{ATLAS:2012pu,Chatrchyan:2011ns,Aaltonen:2008dn,Abazov:2009ac,Chatrchyan:2012bf}.

ATLAS recently reported a 95\% confidence level (CL) exclusion lower limit on
$\Mqst$ of 2.83\TeV as well as a 95\% CL exclusion of a quark contact
interaction with compositeness scale below 7.6\TeV, using 5\fbinv collected at a
center of mass energy of 7\TeV \cite{ATLAS:2012pu}. In these previous studies,
new couplings are always assumed to be equal to the SM gauge couplings
(i.e. $f=\fp=\fs=1$) and the compositeness scale, $\Lambda$, is taken to be
equal to the excited quark mass, $\Mqst$. Gauge and contact interaction
transitions are also typically assumed for excited fermion searches at HERA
where a lower limit on $\Mqst$ of 252\GeV assuming $\fs=0$ was set
\cite{hera}. Finally, production via contact interactions is generally probed in
excited lepton searches at hadron colliders~\cite{ATLAS:2012af,CMS:2012ad},
where excited leptons with masses below 1.9\TeV are excluded for the case where
the contact interaction scale equals the excited lepton mass~\cite{CMS:2012ad}.

Searches for \qStar production use a range of different strategies. The D0
experiment searched for a mass resonance in the Z plus jet system, with the Z
boson detected via its dielectron decay mode \cite{PhysRevD.74.011104}. In this
Letter we search for signs of boosted Z boson decays in the inclusive
\invPtMuMu~spectrum, without specifying the recoiling system so as to be less
model dependent. The choice of \invPtMuMu~as the variable to use in the search
for new physics offers several advantages over \ptMuMu. First, it includes the
coverage of the \ptMuMu~spectrum from a cut-off to infinity without missing any
events. Second, it allows a more natural binning of the spectrum given the
diminishing statistics and worsening resolution with the increasing
\ptMuMu. Most importantly, it turns a broad resonance in the
\ptMuMu~distribution into a narrow peak on top of a rapidly falling background in
\invPtMuMu, thus allowing usage of the methodologies for searches of new
resonances.

\begin{figure}[htb]
\centering
\includegraphics[width=0.25\textwidth]{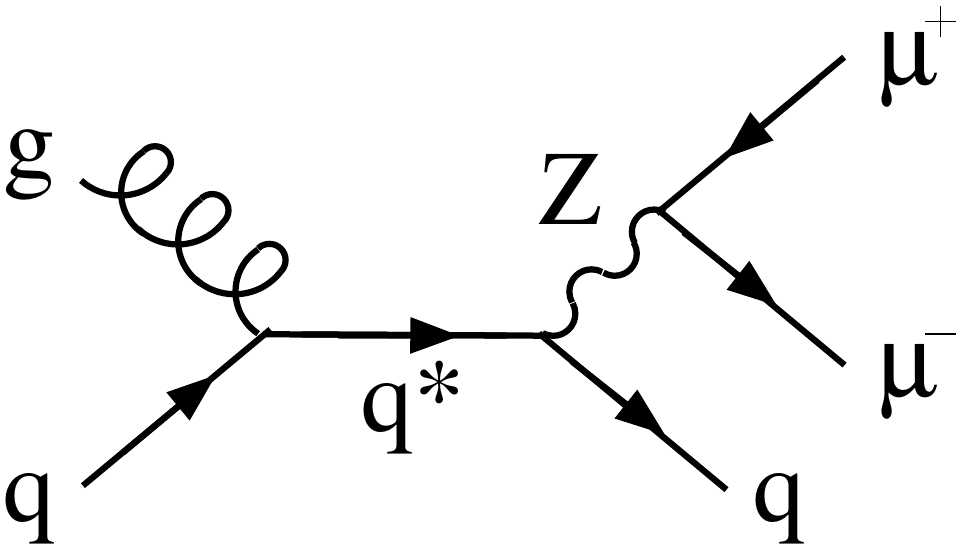}\hspace{0.14\textwidth}
\includegraphics[width=0.25\textwidth]{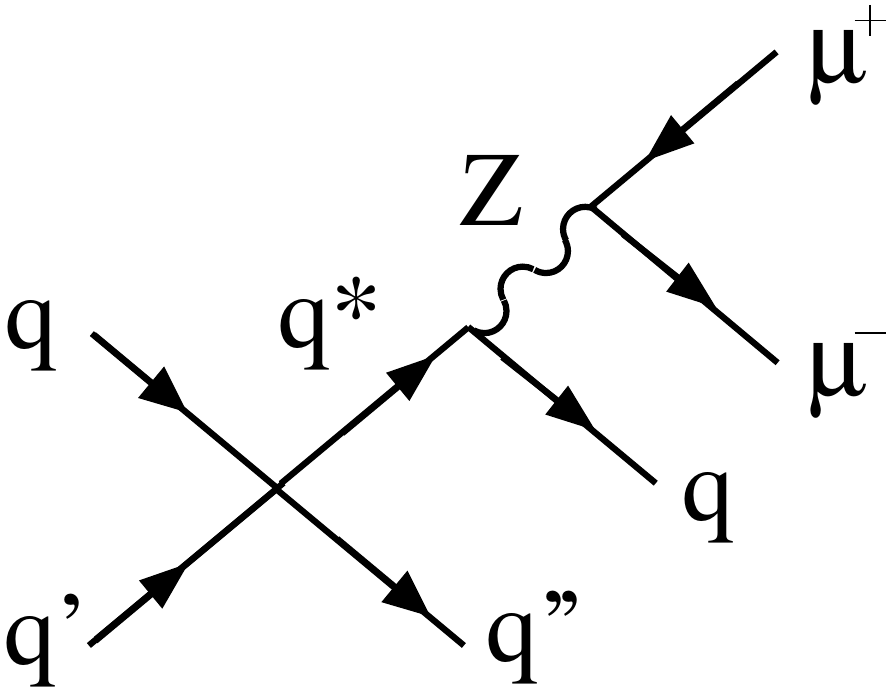}
   \caption{Feynman diagrams for the production of $\cPq^*\to
   \cPq\Z,\:\Z \to \Pgmp \Pgmm$ via (\cmsLeft) a gauge interaction and (\cmsRight) a
   contact interaction. Charge conjugation of an initial state quark is implied
   in both diagrams. Permutation of the quark isospin is implied for the contact interaction
   diagram.}
\label{fig:diagrams}
\end{figure}

The dimuon signature is free of instrumental background, however, it suffers
from the low branching fraction (3.36\%) of the Z decay to $\Pgmp\Pgmm$. The
high luminosity delivered by the LHC collider makes it possible to present
results on this final state for the first time with data recorded at a center of
mass energy of 7\TeV.  In order to reduce the model dependence of the results,
the analysis is not restricted to the common assumptions, $f=\fp=\fs=1$ and
$\Mqst=\Lambda$, but probes a broader phase space. For instance, searches for
new physics in dijet final states do not have sensitivity to models with
$f_s=0$. A reduced parameter space of the excited quark production models is
probed assuming $f=\fp=1$ and using three independent parameters: the mass of
the excited quark, \Mqst, the compositeness scale, $\Lambda$, and the strong
coupling scale factor, $f_s$. Only first generation excited quarks, degenerate
in mass (u$^*$, d$^*$), are considered, as they have the largest production
cross sections in proton collisions. The u$^*$ (d$^*$) branching fraction to
\cPZ~is 3\% (5\%) for gauge only couplings with $f_s=1$, increasing to slightly
over 20\% (30\%) with $f_s=0$ \cite{baur,baur2}. The production of excited
quarks via a gauge transition (qg $\to$ q$^*$) is treated separately from the
production via a contact interaction diagram
($\cPq\cPq^\prime\to\cPq^{\prime\prime}\cPq^*$), though in both cases only the
gauge decay to a Z boson ($\cPq^*\to \cPq\Z$) is considered.
Nevertheless, when production via contact interaction is assumed, the
$\cPq^*\to\cPq\Z$ decay branching fraction is calculated assuming both gauge and
contact interaction mechanisms contribute to the total decay width. The gauge
interaction transitions imply that Z bosons are produced in association with a
quark (qZ), whereas contact interaction production yields Z bosons accompanied
by two quarks (qqZ), as illustrated in Fig.~\ref{fig:diagrams}.

This study uses proton proton collisions data at $\sqrt{s}=7\TeV$, corresponding
to an integrated luminosity of \anaLumi~recorded by the Compact Muon Solenoid
(CMS) experiment at the Large Hadron Collider (LHC). The central feature of the
CMS apparatus is a superconducting solenoid, of 6\unit{m} internal diameter,
providing a field of 3.8\unit{T}. Inside the magnet coil are the silicon pixel
and strip tracker, the lead tungstate crystal electromagnetic calorimeter, and
the brass/scintillator hadron calorimeter. Muons are detected in gas ionization
detectors embedded in the magnet steel return yoke. In addition to the barrel
and endcap detectors, CMS has extensive forward calorimetry.  The trigger
system, composed of a custom hardware layer feeding into a commercial processor
farm (High Level Trigger, HLT), reduces the event rate to approximately 300~Hz
for storage and further analysis. A detailed description of the CMS apparatus
may be found elsewhere~\cite{CMS:2008zzk}.

Anomalous production of Z bosons arising from heavy new particles and decaying
into dimuon final states is characterized by a pair of oppositely charged
isolated muons with an invariant mass consistent with that of the Z boson and a
high dimuon transverse momentum. The dominant irreducible background is due to
SM $\Z/\gamma^* \to \Pgmp\Pgmm$ production. The other background sources
considered in the analysis are from prompt muon processes. Prompt muons are
defined to be muons originating directly at the primary vertex or originating
from the decays of short lived particles. The background sources of prompt muons
considered are $\ttbar$, diboson (WW, WZ, ZZ), and $\Z/\gamma^* \to \Pgt^+\Pgt^-
\to \Pgmp \Pgmm$+ X production. In addition, jets may be misidentified as muons
and contribute to the dimuon transverse momentum spectrum through multijet and W
plus jets final states (non--prompt muon backgrounds). The backgrounds were
modeled using simulated samples produced with the full \GEANTfour~\cite{GEANT4}
based CMS detector simulation, trigger emulation, and event reconstruction
chain. Different samples of SM Z boson production were generated with \POWHEG
v1.1 \cite{Alioli:2008gx, Nason:2004rx, Frixione:2007vw} and the \MADGRAPH
matrix element generator
\cite{MADGRAPH}, both interfaced to the \PYTHIA v6.424
\cite{Sjostrand:2006za} parton shower generator. Diboson and QCD processes with
a muon in the final state were modeled using \PYTHIA. Events from $\ttbar$ and W
plus jets were modeled using \MADGRAPH interfaced to the \PYTHIA parton shower
generator. For the excited quark modeling, we relied on simulations assuming
separately either gauge or contact interaction production, generated for mass
points ranging from $\Mqst=500$ to $\Mqst=2000$\GeV for the gauge interaction
and from $\Mqst=500$ to $\Mqst=2300$\GeV for the contact interaction production,
both in steps of 100\GeV and using \PYTHIA. Gauge decays to a Z boson are assumed
for both production choices. Signal samples are based on the leading order (LO)
compositeness model described in Ref.~\cite{baur,baur2}, obtained with the
CTEQ6L1~\cite{Nadolsky:2008zw} parametrization for the parton distribution
functions and the Z2 underlying event tune~\cite{Field:2010bc}. 

Events are selected offline to have two high \pt ($\pt>45\GeV$), oppositely
charged, isolated muons. Events used in the analysis were collected using a
single muon trigger. The algorithm requires a muon candidate to be found in the
muon detectors by the first level trigger system. The candidate track is then
matched to a silicon tracker track, forming a HLT muon. The HLT muon is required
to exceed a \pt threshold of 40\GeV and to be reconstructed in $|\eta|<$ 2.1,
where the pseudorapidity is defined as $\eta=-\ln[\tan(\theta/2)]$, where
$\theta$ is the polar angle with respect to the direction of the
counterclockwise beam. A right handed coordinate system is used in CMS, with the
origin at the nominal collision point, the $x$ axis pointing to the center of
the LHC ring, the $y$ axis pointing up (perpendicular to the LHC plane), and the
$z$ axis along the anticlockwise beam direction. The azimuthal angle $\phi$ is
the angle relative to the positive $x$ axis measured in the $x$-$y$ plane.

Muon candidates are reconstructed offline with two algorithms whose performance
and validations are discussed in Ref.~\cite{Chatrchyan:2012xi}.  In the first
algorithm, known as ``tracker muon'', tracks are fit with hits in the silicon
tracker, propagated outward, and matched to hits in the muon system. In the
second algorithm, known as ``global muon'', a global fit is performed to hits
both in the silicon tracker and the muon system. At least one muon candidate is
required to be successfully reconstructed by both algorithms. Requiring both
muons to be reconstructed as global muons would eliminate some background, but
would introduce a 35\% efficiency loss associated with muon pairs with small
opening angles, $\Delta R
\equiv
\sqrt{\Delta \phi^2 + \Delta \eta^2} < 0.3$, which is typical of highly boosted
Z candidates ($\ptMuMu > 600\GeV$). The track associated with each muon
candidate is required to have hits in at least 8 layers of the silicon tracker,
at least one hit in the pixel detector, and a magnitude of the transverse impact
parameter below 0.2\unit{cm} to be consistent with a particle emanating from the
primary interaction vertex. The candidate reconstructed as a global muon is
further required to have hits in at least two different muon detector
stations~\cite{Chatrchyan:2012xi}. As the muon passes through the steel of the
magnet return yoke, multiple scattering and radiative processes can alter the
muon trajectory. To further improve the muon momentum resolution at high \pt,
CMS developed a specialized offline reconstruction algorithm to measure the
single muon transverse momentum, within the global fit reconstruction algorithm,
called ``Tune P''~\cite{Chatrchyan:2012xi}, which has been employed in this
analysis. This algorithm has been shown to give a high efficiency for muons
($\simeq$95\%), which is independent of their transverse momentum up to values
of a few hundred GeV. The momentum vector of the offline muon candidate
reconstructed with the global fit algorithm must be matched in direction to the
HLT candidate that triggered the event. If both muons are reconstructed with the
global fit algorithm, only one of them is required to match an HLT muon
satisfying the trigger requirements. Both muons reconstructed offline must have
$\pt> 45\GeV$ and $|\eta|< 2.1$. The trigger efficiency for a single muon that
passes all offline selection criteria is 92\%. Muon candidates from the same
vertex are selected by performing a common vertex fit and requiring the vertex
$\chi^2$ to be below 10.  The muon pair is required to have an invariant mass
consistent with the Z boson, \massCut.

Finally, events are required to contain at least one reconstructed primary
interaction vertex with at least four tracks, located within 2\unit{cm} of the
center of the detector in the direction transverse to the beam and within
24\unit{cm} in the direction along the beam. These requirements help to reject
cosmic-rays background. Additional suppression of muons from cosmic-rays is
obtained by requiring the three dimensional opening angle between the two muons
to be smaller than $\pi - 0.02$ radians. These requirements combined with the
offline selection criteria bring down the cosmic-rays background contamination
to a negligible level.

For the production processes considered here, simulations show that the muons
from Z decay are usually isolated from the hadronic activity in the event. To
measure the isolation for each muon candidate, a cone is constructed of radius
$R = 0.3$ around the track direction at the primary interaction
vertex. We estimate the activity around the lepton by computing the scalar sum
of the transverse momenta of charged tracks within this cone, excluding the muon
candidate itself. Only tracks with distance of closest approach to the primary
vertex less than 0.2\unit{cm} along the beam direction are considered. The ratio of
this isolation quantity to the \pt of the muon itself is required to be less
than 0.1. The efficiency of this requirement for dimuon events in the
\cPZ\ mass window is found to be independent of the number of reconstructed
primary vertices, showing that the isolation is insensitive to the effects of
multiple pp interactions (pileup), which is on average nine per bunch
crossing. The simulated samples are reweighted such that the distributions of
the number of reconstructed primary vertices correspond to the measured
distributions in data. The isolation efficiency is strongly dependent on the
\ptMuMu: For highly boosted Z boson decays, the two muons can approach
collinearity, and therefore can affect each other's isolation calculation. We
correct the isolation measurement of a muon by excluding the other muon
candidate from the isolation sum if the two are closer than $\Delta R = 0.3$ and
if the scalar sum of transverse momenta defining the isolation before the
correction is applied is greater than 90\% of the transverse momentum of the
other muon. The correction recovers the inefficiency induced by the isolation
algorithm, and a flat isolation efficiency as a function of \ptMuMu is obtained.

\begin{figure}[hbt]
\begin{center}
\includegraphics[width=\cmsFigWidth]{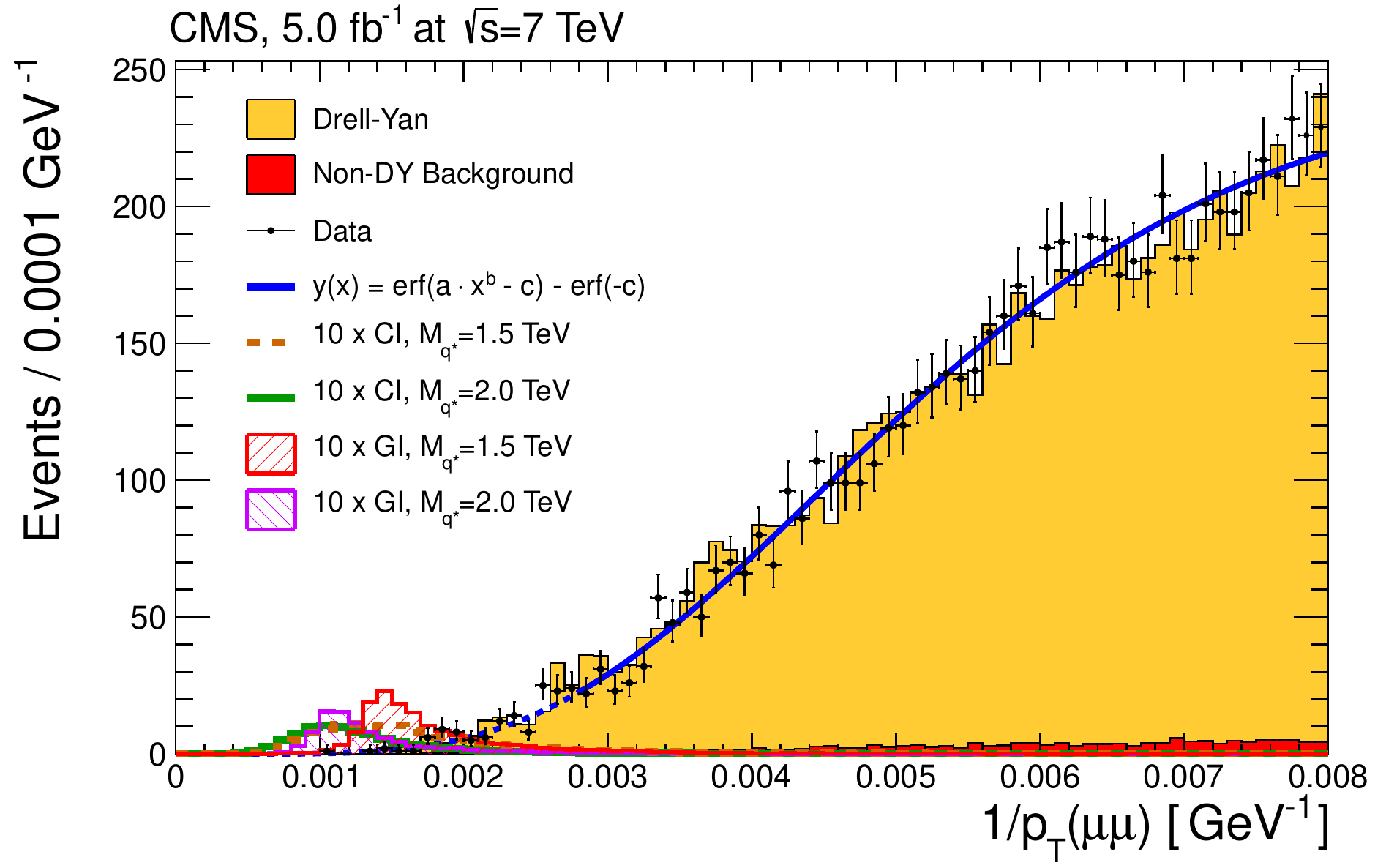}
\includegraphics[width=\cmsFigWidth]{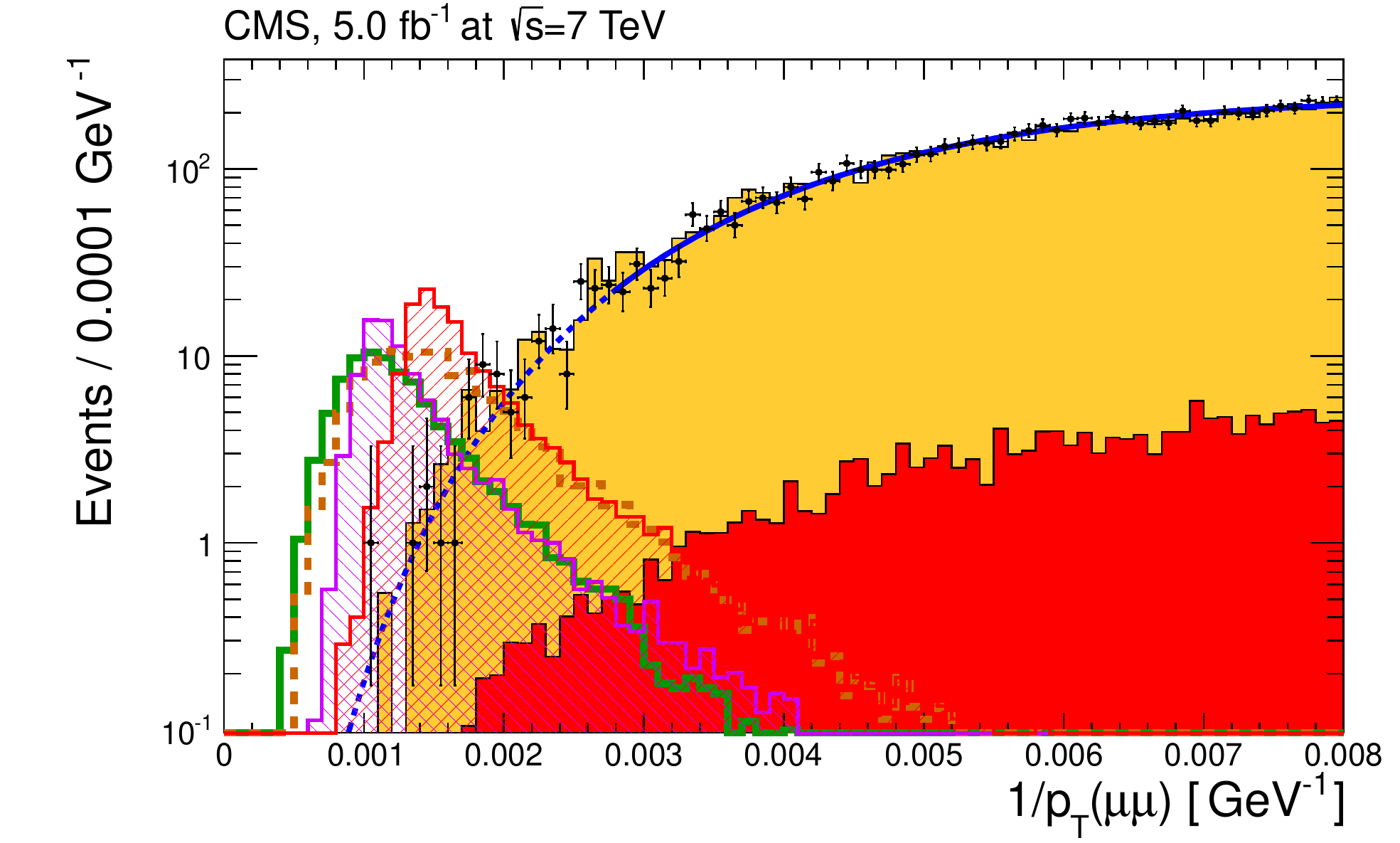}
\caption{Distributions of \invPtMuMu~for data (points with error bars), simulated
         SM backgrounds (stacked histograms), and simulated signal models
         (overlaid histograms). Top: linear scale. Bottom: logarithmic scale.
         The signal models use either gauge interaction (GI) or contact
         interaction (CI) simulated with the assumptions of $f=\fp=\fs=1$ and
         $\Mqst=\Lambda$. The blue solid line corresponds to the analytical
         template fit in the region defined as
         \invPtMuMu~$\in$~[0.0028,~0.008]\invGeV. The contribution labeled
         ``Non--DY Background'' represents the sum over all the other sources of
         prompt backgrounds, \ttbar, $\Z \to \Pgt^+\Pgt^-$, diboson, and
         non--prompt backgrounds, jets misidentified as muons through multijet
         and W plus jets final states.  The total number of SM background events
         is rescaled to the number of events observed in data, using the
         relative background contributions as obtained from simulation. The
         normalization of the signal distributions is increased by a factor ten.
         \label{fig:zInvPt} }
\end{center}
\end{figure}

The distribution of the inverse of the observed dimuon transverse momentum
(\invPtMuMu) is shown in Fig.~\ref{fig:zInvPt} for candidates passing all the
selection criteria and having \invPtMuMu~less than
0.008\invGeV~($\pt(\Pgm\Pgm)>125\GeV$). There are 7044 (29) events observed for
\invPtMuMu less than 0.008 (0.002)\invGeV. The dimuon candidate with the highest
transverse momentum is found to have $\ptMuMu= 940\GeV$.
Simulations predict all background components other than $\Z/\gamma^* \to \Pgmp
\Pgmm$ production to contribute from less than 2\% for \invPtMuMu $<$ 0.008
\invGeV up to about 2.5\% for \invPtMuMu $<$ 0.002 \invGeV, resulting in a negligible
impact on the overall shape of the \invPtMuMu spectrum. The simulation
predictions were cross checked by estimating the prompt and non--prompt muon
background contributions from the data.  The dominant non Drell--Yan background
contributions at high dimuon transverse momentum arise from $\Z/\gamma^* \to
\Pgt^+ \Pgt^-$, $\ttbar$ and diboson production. All these processes are flavor
symmetric and produce twice as many $\Pe\Pgm$ pairs as $\Pe\Pe$ or $\Pgm\Pgm$
pairs. The prompt muon non Drell--Yan background is also estimated by comparing
the transverse momentum spectrum of $\Pe^\pm\Pgm^\mp$ events between data and
simulation and correcting for differences in the geometric acceptances and
efficiencies. This method predicts 46 ($\ll$1) isolated $\Pgm^\pm \Pgm^\mp$
pairs with $\invPtMuMu<0.008\ (0.002)\invGeV$. The estimate of the residual
contribution from background events with at least one non--prompt or
misidentified muon was made by looking at events selected from the data sample
with single muons that pass all selection cuts except the isolation
requirement. A probability map is created for a muon to pass the isolation
criteria employed in the analysis as a function of \pt and $\eta$. This
probability map is corrected for the expected contribution from events with
single prompt muons from $\ttbar$ and W decays and for the observed correlation
between the probabilities for two muons in the same event. The probability map
is used to predict the number of background events with two isolated muons based
on the sample of events that have two non--isolated muons. This procedure, which
has been validated using simulated events, predicts a background of less than
one event for $\invPtMuMu<0.008\invGeV$. More details on similar techniques are
described in Ref.~\cite{zprimePaper}.

The main background contribution arises from SM processes producing $\Z/\gamma^*
\to \Pgmp \Pgmm$, and it is evaluated from a template fit to the
distribution observed in data. Both simulations and
data driven background estimations showed that all the sources of background
other than the $\Z/\gamma^* \to \Pgmp \Pgmm$ have negligible impact on the
total background shape template. Therefore, the choice of the analytical
template is then driven by studies with $\Z/\gamma^* \to \Pgmp \Pgmm$
simulated events, which were generated with \POWHEG and \MADGRAPH and hadronised
using \PYTHIA:

\begin{equation}\label{eq:bkgFitOnData}
 y(x) = \erf(a \cdot x^{b} - c) - \erf(-c)\,,
\end{equation}

where $\erf$ is the error function and {$x$} is \invPtMuMu. The parameters
$a$, $b$, and $c$ of the background analytical template are obtained by fitting
the region $\invPtMuMu\in[0.0028,0.008]\invGeV$. This choice avoids potential
contamination from a signal with $\Mqst\ge1.0\TeV$, which could otherwise bias
the fit. This effect has been tested with Monte Carlo pseudo-experiments
generated with a sample statistically equivalent to 5.0\fbinv and based on
analytical forms of the \invPtMuMu spectrum to describe both the background and
the signal shapes.
Conversely, the larger number of events in the higher \invPtMuMu~region provides a
robust estimation of the background, hence potential contamination due to
\qStar signals with $\Mqst<1.0\TeV$ negligibly biases the fit.
Simulated Drell--Yan events were used to demonstrate that the number of events
predicted by the fit to the \invPtMuMu~distribution is not sensitive to the
choice of the \invPtMuMu~fitting region. The resulting fit to the data is shown
in Fig.~\ref{fig:zInvPt}. The fit predicts $7021 \pm 110$ ($19.6 \pm 0.3$)
events for \invPtMuMu less than 0.008 (0.002)\invGeV. The uncertainty
on the estimated number of events includes the uncertainties associated with the
shape predictions as well as the uncertainty on the total event yield from the
fitting procedure.

The observed \invPtMuMu~spectrum shape agrees with expectations based on SM
processes. With no evidence for new physics, we proceed to set 95\% CL upper
limits on the cross section ($\sigma$) for an excited quark production and decay
process $\qStar \to \cPq \Z$. For the calculation of the limits, we adopt the
frequentist construction CL$_{\mathrm{s}}$~\cite{junk-1999-434,cls}. We use an
extended likelihood~\cite{Beringer:1900zz}, built from characteristic signal and
background probability density function templates. In all cases, we use the
\roostats implementation of the algorithm~\cite{roostats}.
In order to parametrize the underlying shape from the excited quark decay model,
we rely on simulations of models assuming only gauge interaction production and
models assuming only contact interaction production. The upper limit on the
cross section for a model of interest is obtained from the upper limit on the
number of signal events divided by the integrated luminosity, $\int
\mathcal{L}\,\rd{t}$, and the detector acceptance times the efficiency, $\acc \times
\epsilon$. The acceptance, $\acc$, is defined as the fraction of generated
dimuon candidates that have invariant mass \massCut~and both muons with
$|\eta|<$2.1 and $\pt> 45\GeV$. For the excited quark model produced with gauge
interaction, the signal acceptance times efficiency after the complete selection
criteria for dimuon
$\invPtMuMu<0.008\invGeV$ is between $42\%$ ($\Mqst = 0.5\TeV$) and $73\%$
($\Mqst = 2.0\TeV$). For the contact interaction production scenario, the
acceptance times efficiency is between $51\%$ ($\Mqst = 0.5\TeV$) and
$73\%$ ($\Mqst = 2.3\TeV$).

The values for the $\acc \times \epsilon$ are obtained from simulation, but the
efficiency of passing the offline selection criteria, $\epsilon$, is corrected
by a factor $S_{\epsilon}$, which accounts for data simulation
discrepancies. The single muon triggering and particle identification
efficiencies are measured in a sample of inclusive $\Z/\gamma^* \to \Pgmp
\Pgmm$ events in data and MC simulation separately using a tag and probe
technique \cite{Khachatryan:2010xn,Chatrchyan:2012xi}. The difference between
the efficiency measured in data and simulation is found to be flat as a function
of \ptMuMu and a single scale factor is used to correct the dimuon selection
efficiency, $\epsilon$.
The final value of $S_{\epsilon}$ used in the analysis is $0.98 \pm 0.03$. Most
of the systematic uncertainty in the scale factor arises from the low statistics
region where $\Delta R$ between the muons is $<$0.3. We assign a systematic
uncertainty of 2\% for the detector acceptance, evaluated by varying the final
and initial state radiation, as well as parton distribution functions (PDF) sets
in the simulation of the signal models. The leading order prediction for the
signal cross section is assumed to have no uncertainty. The PDF uncertainties on
the final selection signal acceptance have been calculated using the
PDF4LHC~\cite{Botje:2011sn} prescription, using the PDF sets
CT~\cite{Nadolsky:2008zw}, MSTW~\cite{Martin:2009iq} and
NNPDF~\cite{Ball:2010de}. The PDF uncertainties are found to be well below 1\%
for the \qStar model considered. The systematic uncertainty in the integrated
luminosity is estimated to be 2.2\%~\cite{lumiPAS}. The background yield is
treated as a nuisance parameter in the limit setting procedure, and its
estimated value is compatible with the previous fit estimation of the background
normalization. Its systematic uncertainty comes directly from the fit to the
data and is estimated to be 2\%.
The uncertainties derived on the shape parameters of the template fit do not
affect the final results. Other fit templates were tried, but gave fits that
tended to over estimate the background yield, resulting in more stringent
exclusion limits. For each systematic uncertainty, we use a log normal
distribution for the nuisance parameters in the likelihood construction.

\begin{figure}[htb]
\centering
\includegraphics[width=\cmsFigWidth]{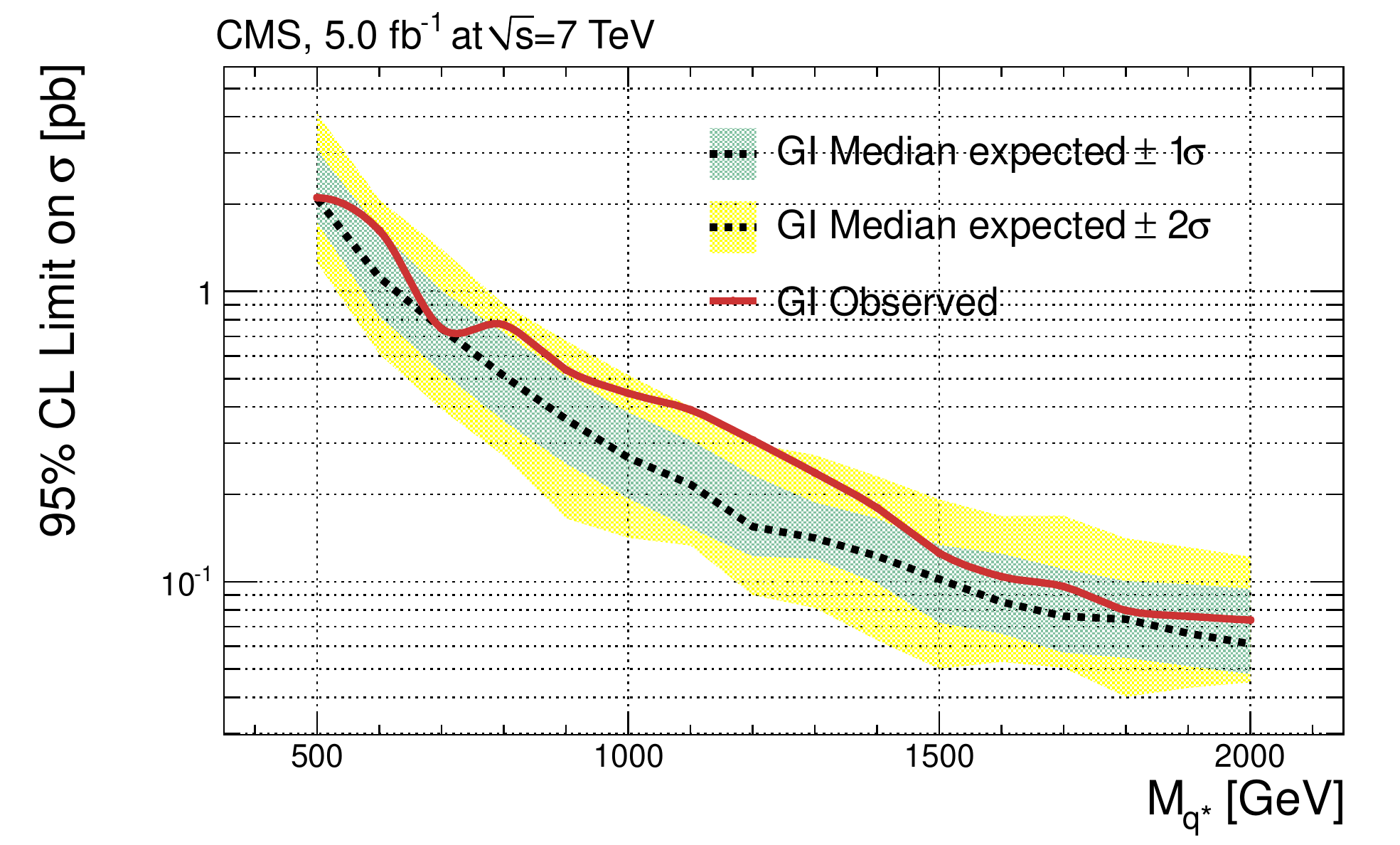}
\includegraphics[width=\cmsFigWidth]{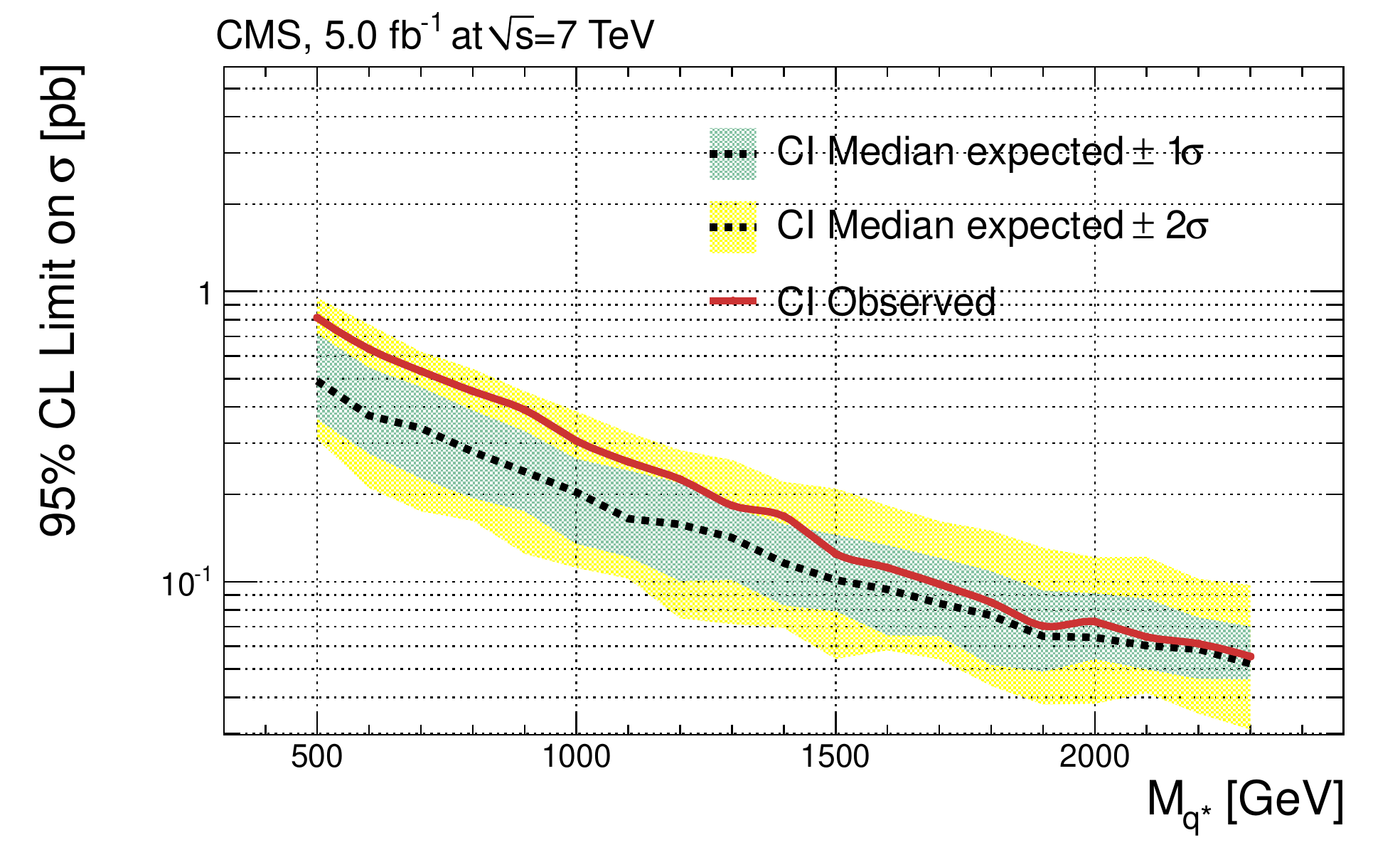}
   \caption{Observed and expected 95\% CL upper limits on the cross section
    of $\qStar \to \cPq\Z$ for the
    benchmark models derived with the CL$_{\mathrm{s}}$ limit setting criterion:
    (top) gauge interaction production, (bottom) contact interaction production.
    The results shown are obtained assuming all new couplings equal to the SM
    couplings between ordinary fermions ($f=\fp=\fs=1$) and
    $\Lambda=\Mqst$.}
   \label{fig:limits}
\end{figure}

The 95\% CL upper limits on cross section for both gauge interaction and contact
interaction are shown in Fig.~\ref{fig:limits}.  The expected limits and the
one  and two standard deviation bands are overlaid on top of the observed
limit.
Generally, the limits are within the two standard deviation bands. A small
excess in the number of observed events in data over the fit predictions, at the
level of two standard deviations, is found in the region $\invPtMuMu <
0.002\invGeV$, resulting in limits that are less stringent than expected for the
gauge interaction models with $1.0 < \Mqst < 1.4 \TeV$.
Since the gauge interaction yields a two body final state while the contact
interaction production yields a three body final state, the gauge interaction
signal distributions are generally narrower than the contact interaction
distributions. Owing to this and the sharp left edge of the gauge interaction
distributions (Fig.~\ref{fig:zInvPt}), a small excess in this region is
sufficient for favoring the signal hypothesis. The limits at the $\qStar$~mass
value probed are correlated as a result of the large width of the signal shapes
present in the \invPtMuMu spectrum.

\begin{figure}[htb]
\begin{center}
\includegraphics[width=\cmsFigWidth]{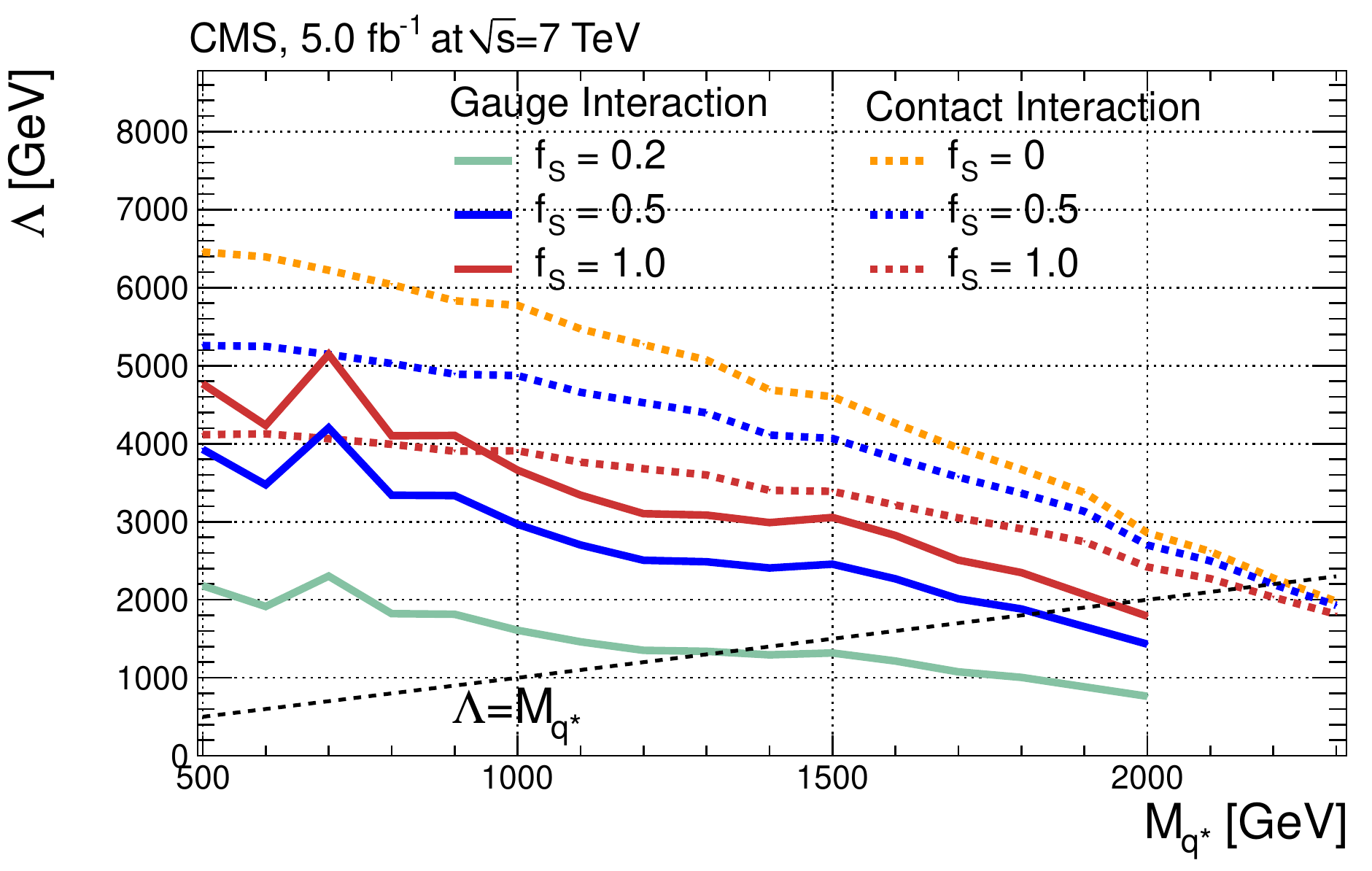}
   \caption{Contours of excluded parameters for the gauge and contact interaction \qStar~production mechanisms. For a given
             $\fs$~assumption, each (\Mqst, $\Lambda$) below the limit curve
             is excluded. The bump in the $\Lambda$ contour at $\Mqst=700\GeV$
             for the gauge interaction production models is a direct consequence
             of the downward fluctuation on the upper limit on the cross section
             observed at $\Mqst=700\GeV$ (Figure~\ref{fig:limits}).
             \label{fig:limContours} }
\end{center}
\end{figure}

The limits on cross section are then translated into contours in the parameter
space of the gauge interaction and contact interaction models. For a given pair
of values of \fs and \Mqst, we probe different values of the compositeness
scale $\Lambda$ in steps of 100\GeV and select the value of $\Lambda$ producing
the closest cross section to the observed limit. We use \PYTHIA to compute
the u$^*$ and d$^*$ cross section values at leading order. The excited quark
branching fractions are computed according to the equations reported in
Ref.~\cite{baur}.
In Fig.~\ref{fig:limContours} we report the final limit contours in the
\Mqst and $\Lambda$ plane for different assumptions of the strong coupling,
\fs, separately in gauge interaction and contact interaction production. The
limit contours directly translate the results obtained on the cross section: a
lower value for the upper limit on the cross section corresponds to a higher
value for the lower limit on $\Lambda$; \eg, the downward statistical
fluctuation at $\Mqst=700\GeV$ in the gauge interaction production model
translates into a bump in the corresponding $\Lambda$ contour.

In summary, a search for anomalous production of highly boosted
\cPZ~bosons in the dimuon decay channel has been performed using proton proton
collision data with an integrated luminosity of \anaLumi~collected at
$\sqrt{s}=7\TeV$ by the CMS experiment. The \cPZ~transverse momentum
distribution observed is consistent with SM expectations. Limits are derived on
the specific model of excited quark production and decay in $\qStar \to
\Pq\Z$. We report 95\% exclusion contours in the compositeness scale versus
excited quark mass plane for two production scenarios and for several choices of
the relative coupling to gluons.  Under the assumptions for the parameters
$\Mqst=\Lambda$ and $f=\fp=\fs=1$, our limits exclude excited quarks at 95\% CL
with $\Mqst<\limitsGIfsOne$~for gauge production and $\Mqst<\limitsCIfsOne$~for
the contact interaction respectively.  In comparison, the corresponding best
exclusion limits for the gauge production of \qStar from a search in the dijet
final state is $\Mqst< 2.83\TeV$~\cite{ATLAS:2012pu}. Nevertheless, the results
from this analysis probe a complementary \qStar electroweak decay and extend the
limits to regions where the default assumptions on $\Mqst = \Lambda$ and
$\fs=1$ have been relaxed. For \qStar created through the contact interaction
scenario, we exclude a large section of the strong coupling phase space,
including $\fs=0$ with a mass up to
\limitsCIfsZero, much higher than the $\fs=0$ limit of 252\GeV set by
H1~\cite{hera} in ep collisions.

We congratulate our colleagues in the CERN accelerator departments for the
excellent performance of the LHC machine. We thank the technical and
administrative staff at CERN and other CMS institutes, and acknowledge support
from BMWF and FWF (Austria); FNRS and FWO (Belgium); CNPq, CAPES, FAPERJ, and
FAPESP (Brazil); MEYS (Bulgaria); CERN; CAS, MoST, and NSFC (China); COLCIENCIAS
(Colombia); MSES (Croatia); RPF (Cyprus); MoER, SF0690030s09 and ERDF (Estonia);
Academy of Finland, MEC, and HIP (Finland); CEA and CNRS/IN2P3 (France); BMBF,
DFG, and HGF (Germany); GSRT (Greece); OTKA and NKTH (Hungary); DAE and DST
(India); IPM (Iran); SFI (Ireland); INFN (Italy); NRF and WCU (Korea); LAS
(Lithuania); CINVESTAV, CONACYT, SEP, and UASLP-FAI (Mexico); MSI (New Zealand);
PAEC (Pakistan); MSHE and NSC (Poland); FCT (Portugal); JINR (Armenia, Belarus,
Georgia, Ukraine, Uzbekistan); MON, RosAtom, RAS and RFBR (Russia); MSTD
(Serbia); SEIDI and CPAN (Spain); Swiss Funding Agencies (Switzerland); NSC
(Taipei); ThEP, IPST and NECTEC (Thailand); TUBITAK and TAEK (Turkey); NASU
(Ukraine); STFC (United Kingdom); DOE and NSF (USA).

\bibliography{auto_generated}   

\cleardoublepage \appendix\section{The CMS Collaboration \label{app:collab}}\begin{sloppypar}\hyphenpenalty=5000\widowpenalty=500\clubpenalty=5000\textbf{Yerevan Physics Institute,  Yerevan,  Armenia}\\*[0pt]
S.~Chatrchyan, V.~Khachatryan, A.M.~Sirunyan, A.~Tumasyan
\vskip\cmsinstskip
\textbf{Institut f\"{u}r Hochenergiephysik der OeAW,  Wien,  Austria}\\*[0pt]
W.~Adam, E.~Aguilo, T.~Bergauer, M.~Dragicevic, J.~Er\"{o}, C.~Fabjan\cmsAuthorMark{1}, M.~Friedl, R.~Fr\"{u}hwirth\cmsAuthorMark{1}, V.M.~Ghete, J.~Hammer, N.~H\"{o}rmann, J.~Hrubec, M.~Jeitler\cmsAuthorMark{1}, W.~Kiesenhofer, V.~Kn\"{u}nz, M.~Krammer\cmsAuthorMark{1}, I.~Kr\"{a}tschmer, D.~Liko, I.~Mikulec, M.~Pernicka$^{\textrm{\dag}}$, B.~Rahbaran, C.~Rohringer, H.~Rohringer, R.~Sch\"{o}fbeck, J.~Strauss, A.~Taurok, W.~Waltenberger, G.~Walzel, E.~Widl, C.-E.~Wulz\cmsAuthorMark{1}
% ...
\vskip\cmsinstskip
\dag:~Deceased\\
1:~~Also at Vienna University of Technology, Vienna, Austria\\
% ...

\end{sloppypar}
\end{document}